\documentclass[12pt,a4paper]{article}

\usepackage[T1]{fontenc} 
\usepackage{amsmath}
\usepackage{amsfonts}
\usepackage{amssymb}
\usepackage{graphicx}
\usepackage{subfig}
\usepackage{hyperref}
\def\be{\begin{equation}}
\def\ee{\end{equation}}

\newcommand{\de}{\partial}
\def\beq{\begin{equation}}
\def\eeq{\end{equation}}

\newcommand{\Gam}{\Gamma}
\newcommand{\del}{\delta}

\newcommand{\eps}{\varepsilon}

\newcommand{\x}{\vec{x}}
\newcommand{\rmd}{\textrm{d}}
\newcommand{\nab}{\nabla}

\newcommand{\Mpl}{M_{\textrm{Pl}}}
\newcommand{\fnl}{f_{\textrm{NL}}}
\newcommand{\avg}[1]{\langle #1 \rangle}
\renewcommand{\k}{\vec{k}}
\newcommand{\R}{\mathcal{R}}

\begin{document}
\def\thefootnote{\fnsymbol{footnote}}

\begin{center}
\Large{\textbf{Galilean symmetry in the effective theory of inflation:\\
new shapes of non-Gaussianity}} \\[0.5cm]
 
\large{Paolo Creminelli$^{\rm a}$, Guido D'Amico$^{\rm b,c,d}$, Marcello Musso$^{\rm a}$, 
 \\[.1cm] Jorge Nore\~na$^{\rm b,c,e}$ and Enrico Trincherini$^{\rm b,c}$}
\\[0.5cm]

\small{
\textit{$^{\rm a}$ Abdus Salam International Centre for Theoretical Physics\\ Strada Costiera 11, 34151, Trieste, Italy}}

\vspace{.2cm}

\small{
\textit{$^{\rm b}$ SISSA, via Bonomea 265, 34136, Trieste, Italy}}

\vspace{.2cm}

\small{
\textit{$^{\rm c}$ INFN - Sezione di Trieste, 34151 Trieste, Italy}}

\vspace{.2cm}

\small{
\textit{$^{\rm d}$ Center for Cosmology and Particle Physics \\
Department of Physics, New York University \\
4 Washington Place, New York, NY 10003, USA}}

\vspace{.2cm}

\small{
\textit{$^{\rm e}$ Institut de Ci\`encies del Cosmos (ICC), \\
Universitat de Barcelona, Mart\'i Franqu\`es 1, E08028- Spain}}

\vspace{.2cm}

\end{center}

\vspace{.8cm}

\hrule \vspace{0.3cm}
\noindent \small{\textbf{Abstract}\\ 
We study the consequences of imposing an approximate Galilean symmetry on the Effective Theory of Inflation, the theory of small perturbations around the inflationary background. This approach allows us to study the effect of operators with two derivatives on each field, which can be the leading interactions due to non-renormalization properties of the Galilean Lagrangian. In this case cubic non-Gaussianities are given by three independent operators, containing up to six derivatives, two with a shape close to equilateral and one peaking on flattened isosceles triangles. The four-point function is larger than in models with small speed of sound and potentially observable with the Planck satellite.}
\\
\noindent
\hrule
\def\thefootnote{\arabic{footnote}}
\setcounter{footnote}{0}

\section{Introduction}
In this paper we study the observational consequences of the presence of an approximate Galilean symmetry \cite{Nicolis:2008in} in the theory of inflation, or more precisely in the theory describing the perturbations around an inflating solution. While it is always important to study all the possible symmetries that can constrain the dynamics of a physical system, this is particularly true for single field inflation since, given its simplicity, there are not many symmetries one can impose. In a recent paper \cite{Burrage:2010cu} (see also \cite{Kobayashi:2010cm,Mizuno:2010ag}) Burrage {\em et al.} studied inflaton models endowed with the Galilean symmetry
\be
\phi \to \phi + b_\mu x^\mu + c \;,
\ee
with $b_\mu$ and $c$ constant. They restricted their analysis to Galilean operators of the schematic form $(\partial\phi)^2(\partial^2\phi)^n$, $n \leq 3$, which are the ones that give second order equations of motion of the form $(\partial^2\phi)^{n+1}$  \cite{Nicolis:2008in} . In this way perturbations remain well behaved, even when these operators are important for the background solution, which is the most interesting regime, quite different from standard slow-roll. Interestingly this model can give rise to large non-Gaussianities and, even more interestingly, the cubic operators which generate them contain four derivatives, so that they are naively distinguishable from the ones generated in more standard models, where large non-Gaussianities are linked to a low speed of sound and the cubic operators have three derivatives \cite{Alishahiha:2004eh,Chen:2006nt,Cheung:2007st}. Unfortunately, all the operators with four derivatives can be rewritten, as we will explain below, in terms of the ones with three using the linear equations of motion; this implies that this class of models does not give rise to a distinctive shape of non-Gaussianity. This fact motivates us to look beyond the Galilean operators studied in \cite{Burrage:2010cu}, in particular at the ones with two derivatives on each field. We will see that this is consistent -- Galilean operators with less derivatives are not radiatively generated -- and it gives new shapes for non-Gaussianity, with a maximum of six derivatives.

At first the idea of studying inflationary models where operators of the form $(\partial^2\phi)^n$ are important seems silly. First of all there is an infinite number of operators of this form, with all the possible contractions of indices. Moreover when these operators become important for the unperturbed solution -- which is mandatory if we want them to be relevant for the dynamics of the perturbations --  the equations of motion are of higher order and describe also ghost excitations. In other words the solution we are looking seems outside the regime of validity of the effective field theory (EFT). However this comes from a wrong approach to the problem. What matters is the existence of a good EFT describing fluctuations around a quasi de-Sitter background: this is what we call inflation and this is what is relevant for observations.  This is the approach advocated in \cite{Cheung:2007st} (see also \cite{Creminelli:2006xe}) and we will review it in this context in Section \ref{sec:EFT}. The construction of the theory and its radiative stability are discussed in Section \ref{sec:action} and Appendix \ref{app:NR}. The study of the three-point function is done in Sections \ref{sec:perturbations} and \ref{sec:shapes}, while Section \ref{sec:NG4} is devoted to the four-point function. Conclusions are drawn in Section \ref{sec:conclusions}.


\section{\label{sec:EFT}The (Galilean) effective theory of inflation}


Inflationary observables -- the $n$-point functions of the conserved curvature perturbation on comoving slices, which we call $\zeta$ (\footnote{This quantity is usually called $\R$. However, we follow the notation of~\cite{Maldacena:2002vr}, which has become common in the literature.}) -- are calculated within the theory describing small fluctuations around a quasi de Sitter background. As in any effective theory, we need the cut-off $\Lambda$ to be much larger than the typical energy scale of the problem, i.e.~$H$, which is the scale of quantum fluctuations induced by the cosmological background. This is all we need and what we call effective approach to inflation \cite{Cheung:2007st,Creminelli:2006xe}. 

However one may be more ambitious and require the validity of the theory to be much broader than what is needed to reproduce cosmological observations. Although we experimentally probe small perturbations
\be
\phi(t+\pi(t, \vec x))\;,\qquad H \pi = -\zeta \sim 10^{-5}\;,
\ee
one may wonder whether the regime of validity of the theory encompasses very different backgrounds: for example whether it can describe a solution that starts with zero velocity $\dot\phi =0 $ or that, after the end of inflation, oscillates around a local minimum. In other words whether the theory makes sense also for $\pi \sim t$. This is often implicitly assumed in the standard approach to inflation, where one starts from a Lagrangian, finds a classical inflating solution and then studies small perturbations around it. We stress that, although it is nice to have a single EFT describing quite different classical solutions, that can be used not only to derive the inflationary observables but also, for instance, to address the issue of initial conditions and reheating, this is not required and it may represent an unjustified restriction on the theory of inflation. For large values of $\pi$ the EFT may cease to make sense and the appropriate description may be in terms of completely different degrees of freedom.

 Usually the issue is not appreciated as in the simplest model of inflation, obtained expanding around a classical slow-roll solution of a Lagrangian for a single scalar with minimal kinetic term and potential, one can indeed consistently describe very different backgrounds, assuming a sufficiently heavy UV completion.  
 Things are a bit subtler for inflationary theories with a small speed of sound for perturbations. Although they can be described in the effective approach \cite{Cheung:2007st} in a simple way, these theories can be also explicitly derived from a Lagrangian of the form $P(X,\phi)$, $X \equiv - (\partial\phi)^2$. If $P$ is given, one can trust solutions quite different from the inflating one, at least as long as the kinetic term of perturbations remains healthy \cite{derivatively}. This of course is possible only at a price: while the theory of perturbations around the inflating solution just depends on few parameters, one must know the whole function $P(X,\phi)$ from some UV input if interested in very different backgrounds. 
 
 In the case under study, however, one is forced to give up the (unnecessary) ambition of describing solutions which are very far from the inflating one. We will be interested in a Lagrangian for perturbations of the schematic form
\be
\Mpl^2 \dot H(\partial\pi)^2  + M (\partial^2\pi)^3 + \ldots
\ee
and particularly in the regime of large non-Gaussianities, when the second term, evaluated at $\partial \sim H$ and $H \pi \sim 10^{-5}$, gives a non negligible correction to the kinetic term, say of order $10^{-3}$ to be around the present experimental limits. If we now try to describe solutions with $\partial \sim H$ and $\pi \sim t$ (which implies $H \pi \gg 1$ as $H \sim \eps/t$ in terms of the slow-roll parameter $\eps \equiv -\dot H/H^2$) the importance of the cubic term compared to the quadratic will increase by a factor $10^{5}/\eps$, boosted by the large classical occupation number of $\pi$. In the EFT approach however, higher derivative terms {\em must} be small corrections and {\em must} be evaluated using the lowest order equations of motion (see for example \cite{Simon:1990ic,Simon:1990jn}). If on the other hand higher derivative terms are not small, one should solve the complete equations of motion which would, in this case, contain higher time derivatives and need additional initial conditions. These new degrees of freedom are ghost-like, a clear sign that we are out of the regime of validity of the EFT we started with.


What are the rules to build the effective theory of inflation? In \cite{Cheung:2007st,Creminelli:2006xe}, the emphasis was put on a geometrical approach. Here this would not be very useful, because it is difficult to construct Galilean operators in the geometrical language\footnote{For example one of the key point of the geometrical construction is that, given that there is a privileged slicing of the spacetime, one can project all tensors parallely or perpendicularly to it. However these projectors, once expressed in terms of the scalar degree of freedom, are not Galilean invariant as they contain a single derivative.}. Moreover we will see that the mixing with gravity is negligible so that one is solely interested in the action for $\pi$. The straightforward approach is to write down all the possible $\pi$ operators which are Galilean invariant and realize nonlinearly the Lorentz symmetry. This last requirement is equivalent to build Lorentz invariant operators in terms of $\psi \equiv t + \pi(t,\vec x)$, as $\psi$ linearly realizes the Lorentz symmetry. Once the possible operators at each order are written in terms of $\pi$ the size of each term can be estimated by naturalness arguments. 


\section{\label{sec:action}Building up the action}


The purpose of this paper is to study the theory of perturbations about an inflating background when they are endowed with the Galilean symmetry $\pi \to \pi + b_\mu x^\mu + c$; in particular we are interested in seeing whether we can generate new shapes for the three-point function. Galilean invariance implies that there are at least two derivatives for each $\pi$ in the equation of motion, which means that a cubic term in the action contains at least four derivatives (we are assuming that we can take the decoupling limit and just look at the action for $\pi$, neglecting the mixing with gravity; we will come back to this point later). One has cubic terms with only four derivatives if the action for $\pi$ is completed to Lorentz invariant operators for $\psi$ that have the minimum number of derivatives compatible with the Galilean symmetry, that is $(\partial\psi)^2(\partial^2\psi)^n$, because these terms give equations of motion with exactly two derivatives on each $\pi$ \cite{Nicolis:2008in}.

It is easy to see, however, that cubic operators for $\pi$  with four derivatives do not generate new shapes of non-Gaussianity as they can be rewritten in terms of operators with three derivatives, which are the ones typical of models with a reduced speed of sound\footnote{We thank A.~Gruzinov, M.~Mirbabayi, L.~Senatore and M.~Zaldarriaga for pointing this out to us.} \cite{Alishahiha:2004eh,Chen:2006nt,Cheung:2007st}.  Indeed the possible forms of the four-derivative operators are
\be\label{4deriv}
\ddot \pi \dot\pi^2 \quad \nabla^2\pi \dot\pi^2 \quad \dot\pi \nabla\pi \nabla\dot\pi \quad \ddot\pi (\nabla\pi)^2 \quad \nabla^2\pi (\nabla\pi)^2 \;,
\ee
where with $\nabla$ we indicate spatial derivatives\footnote{Reference \cite{Mizuno:2010ag} claims that the shape induced by the fifth operator in \eqref{4deriv} is independent of the shapes given by operators with three derivatives. However it is straightforward to check that it can be written as a linear combination of the shapes induced by $\dot\pi^3$ and $\dot\pi (\nabla\pi)^2$.}.
To show that they can be rewritten as operators with fewer derivatives, we are going to impose the linear equation of motion (at leading order in slow-roll and with generic speed of sound $c_s$) $\ddot\pi + 3 H \dot\pi - c_s^2 \nabla^2\pi/a^2 = 0$ at the operator level; this is equivalent up to third order to perform a field redefinition. Since we want to compute equal-time correlation functions and not S-matrix elements, in principle one should take into account all the field redefinitions~\cite{Maldacena:2002vr}. However it is  easy to realize that the redefinition we need involves derivatives of the field and therefore, when the modes are much longer than the Hubble radius,  it gives contributions of higher order in the slow-roll parameters. 

The first operator in (\ref{4deriv}) is obviously a total derivative, so that a time derivative can be moved to act on the $a^3$ of $\sqrt{-g}$ giving a term with three derivatives. The second is related to the first using the linear equation of motion, apart from terms with fewer derivatives. The third one can be written as $\frac12 \nabla\pi \nabla \dot\pi^2$ and it is thus related to the second integrating by parts. The fourth is the same as the third integrating by parts in time. And finally the fifth is related to the fourth using again the linear equation of motion. Therefore with four derivatives we cannot generate new forms of three-point function and we are hence motivated to consider terms with more derivatives on $\psi$.




\subsection{Galilean action with more derivatives}


Let us assume to start, besides the standard kinetic term, only with operators that have at least two derivatives on each $\pi$ (and therefore on $\psi$), i.e.~setting to zero the lowest derivative Galilean terms $(\partial\psi)^2 (\partial^2\psi)^n$ (\footnote{Notice that this implies that the kinetic term is standard, i.e.~$c_s=1$.}). In flat space this is consistent as loops will not generate them \cite{Luty:2003vm}. However, the Galilean symmetry cannot be defined in curved space, since there is no natural way of defining a constant vector $b_\mu$. Thus one may worry to generate terms containing the Riemann tensor, of the schematic form $(\partial\psi)^2 (\partial^2\psi)^n R$ and in general all terms in which a couple of derivatives is replaced by the curvature tensor. All these would give a contribution to the $\pi$ correlation functions which is of the same order as the one we started with. This is not the case as it is easy to prove. 

Let us start from a 1PI graph for $\pi$ in flat space: this has (at least) two derivatives on each external leg. Now we add an external graviton line\footnote{We do not consider internal graviton lines, as these would be suppressed by powers of $\Mpl$.}. If it couples to an internal line, it will not reduce the number of derivatives acting on each external $\pi$ line. If the graviton is attached to a vertex it can reduce to one the number of derivatives acting on an external $\pi$ line, when $g$ comes from the Christoffel symbol. But all the other external $\pi$ legs have still 2 derivatives. The loop can generate additional external derivatives, but it cannot reduce them, so we see that it is impossible to generate an operator of the form $(\partial\psi)^2 (\partial^2\psi)^n R$, which contains a term with a single external graviton and two $\pi$'s with a single derivative. One would have at least an additional suppression of the curvature: $(\partial\psi)^2 (\partial^2\psi)^n
  R^2$, which makes these terms subleading for the $\pi$ correlation functions. Additional non-renormalization properties are discussed in Appendix \ref{app:NR}. We conclude that, although the Galilean symmetry cannot be defined in curved space, we can consistently reduce to operators which contain two derivatives on each field. 

Let us come to the explicit construction of the $\pi$ action. The $\pi$ operators at any order can be written starting from Lorentz invariant operators for $\psi$. These are traces of the double derivative matrix $\nabla_\mu\nabla_\nu \psi$. In the following we use $\Psi$ to denote this matrix and the square brackets to indicate the trace so that, for example, $[ \Psi \Psi]$ means $\nabla_\mu\nabla_\nu \psi \, \nabla^\mu\nabla^\nu \psi$.
In order to parametrize the theory of perturbations around a given solution we would like to isolate operators that contain terms linear in $\pi$. Those operators change the background and therefore their coefficients are constrained once the background solution is chosen. On the other hand operators that start quadratic in $\pi$ do not affect the background solution for $\psi$ nor for the metric -- as they have vanishing stress-energy tensor for $\pi = 0$ -- so that their coefficients are not constrained \cite{Creminelli:2006xe,Cheung:2007st}. 

If a term is composed by the product of several traces of products of $\Psi$, $[\Psi \Psi \ldots \Psi] \ldots [\Psi \Psi\ldots \Psi]$, then we can subtract to each trace its background value to make the term at least quadratic. If on the contrary an operator is composed by a single trace $[\Psi^n]$ this cannot be done and the operator will start with a linear term. We will show however that  $[\Psi^n]$ can be written in term of operators, whose linear term is easy to isolate. To do this, consider the sum
\be
\label{eq:total}
\sum_p (-1)^p g^{\mu_1 p(\nu_1)} \ldots g^{\mu_n p(\nu_n)} \nabla_{\mu_1}\nabla_{\nu_1} \psi \ldots \nabla_{\mu_n}\nabla_{\nu_n} \psi \;,
\ee
where $p$ is the parity of the permutation. This sum contains also the single trace of order $n$. For $n >4$ eq.~\eqref{eq:total} trivially vanishes, as there are too many indices to antisymmetrize, so that the $[\Psi^n]$ can be rewritten in terms of products of two or more shorter traces. For $n \leq 4$ the sum does not vanish. In flat space it is a total derivative, as it easy to see calculating the equations of motion, which trivially vanish, but in curved space this does not occur since derivatives do not commute. Nevertheless the sum can be written---integrating by parts one of its derivatives, writing the commutator of derivatives in terms of the Riemann tensor and finally expressing the Riemann tensor of de Sitter in terms of the metric---as
\be
\label{eq:totalpart}
H^2 \sum_p (-1)^p g^{\mu_1 p(\nu_1)} \ldots g^{\mu_{n-1} p(\nu_{n-1})} \nabla_{\mu_1}\psi\nabla_{\nu_1} \psi \ldots \nabla_{\mu_n-1}\nabla_{\nu_n-1} \psi  \quad \quad n\le 4 \;,
\ee 
apart from overall combinatorial factors. Notice that these are the minimal Galilean terms, see Appendix A of \cite{Nicolis:2008in}, up to the quartic. Therefore for $n \leq 4$ the single traces can be written in terms of shorter traces and minimal Galilean terms, cubic and quartic in $\psi$ (\footnote{This represents a nice consistency check of our statement before, that generic operators of the form $(\partial\psi)^2 (\partial^2\psi)^n R$ are not generated. As the operator \eqref{eq:totalpart} is Galilean invariant, it will not generate non-Galilean terms, unless paying an additional power of $R$, i.e.~$(\partial\psi)^2 (\partial^2\psi)^n R^2$, consistently with what we said above. See also Appendix \ref{app:NR}. Moreover, as minimal Galilean terms are not renormalized, we are not going to generate the quintic one, which does not come in the process of rewriting the single trace operators.}).
We conclude that it is consistent to study the theory that contains all operators written in terms of $\nabla_\mu\nabla_\nu \psi$, except the single trace operators, plus the minimal Galilean operators of cubic and quartic order. These two will be suppressed by $H^2$ and therefore will give a contribution of the same order of magnitude as the other ones to the $\pi$ correlation functions.

It is straightforward to write the minimal Galilean operators of cubic and quartic order in such a way that they do not contain terms linear in $\pi$. The cubic (DGP-like) term $\Box\psi(\partial\psi)^2$ can be written as $(\Box\psi+3 H)[(\partial\psi)^2+1]$, just redefining the coefficients of the kinetic term and the cosmological constant term, which are anyway present in the Lagrangian. The procedure is slightly subtler for the quartic term and it is useful to rewrite it as \cite{Deffayet:2009wt}
\be
\begin{split}
& (\Box\psi)^2 (\partial\psi)^2 - 2 \Box\psi \partial_\mu\psi \partial_\nu\psi \nabla_\mu\nabla_\nu\psi - (\nabla_\mu \nabla_\nu\psi)^2 (\partial\psi)^2 + 2 \partial_\mu\psi \nabla_\mu \nabla_\nu\psi \nabla_\nu \nabla_\rho\psi \partial_\rho\psi \\
= & (\partial\psi)^2 \left[2 (\Box\psi)^2 - 2 (\nabla_\alpha\nabla_\beta\psi)^2\right] - (\partial\psi)^2 R^{\mu\nu} \partial_\mu\psi \partial_\nu\psi \;.
\end{split}
\ee
As we are in de Sitter the Ricci tensor is proportional to the metric, so that the second operator is just  $(\partial\psi)^2(\partial\psi)^2$. In this form all terms can be treated similarly to what we did with the DGP one.

Now that we have the rules to build the action in terms of $\psi$, we move to study the third order action for $\pi$, which is relevant for the calculation of non-Gaussianity.


\subsection{\label{sec:perturbations}Cubic action for perturbations}


We are primarily interested in the 3-point function, i.e.~in operators of the form $(\partial^2\pi)^3$. Some of these operators will be independent, while others will be related to quadratic operators by the non-linear realization of the Lorentz symmetry, i.e.~once written in terms of $\psi$. In this second case the scale suppressing the cubic operators is related to the one suppressing the quadratic ones. Schematically we will have 
\be
\label{eq:ducks}
\Mpl^2 \dot H (\partial\pi)^2 + M \left[H (\partial^2\pi)^2 + (\partial^2\pi)^3 + \ldots\right] \;,
\ee
where the factor of $H$ inside brackets comes from the fact that $\nabla_\mu\nabla_\nu\psi$ gives $H \delta_{ij}$ on the background.
Going to canonical normalization we have
\be
\label{eq:duckscan}
(\partial\pi_c)^2 + \frac{1}{\Lambda^2}(\partial^2\pi_c)^2 + \frac{1}{\tilde\Lambda^5}(\partial^2\pi_c)^3 + \ldots
\ee
where
\be
\Lambda^2 = \frac{\Mpl^2 \dot H}{H M}
\ee
and 
\be
\tilde\Lambda^5 = \Lambda^5 \frac{H \Mpl \dot H^{1/2}}{\Lambda^3} \simeq \Lambda^5 \frac{H^3}{\Lambda^3} \cdot \sqrt{\eps} \frac{\Mpl}{H} \;,
\ee
where the separation between the scales is stable only when $\tilde\Lambda > \Lambda$.
Notice that $(\partial^2\pi)^2$ must be treated as a small perturbation to the standard kinetic term with $c_s=1$; therefore we have the usual normalization $H/(\sqrt{\eps}\Mpl) \simeq 10^{-5}$. Cubic non-Gaussianities will be of order $(H/\Lambda)^5$, so that observable non-Gaussianity -- i.e.~larger than $10^{-5}$ corresponding, in the usual parametrization, to $f_{\rm NL} \gtrsim 1$ -- implies $\tilde\Lambda \gg \Lambda$. We conclude that, assuming all independent operators are suppressed by a common scale $\Lambda$, cubic operators that are linked by symmetry to quadratic ones are suppressed by a much higher scale and can be neglected with respect to the independent ones. It is straightforward to see that this holds in general when an operator $(\partial^2\pi)^n$ is related by the Lorentz symmetry to a lower dimensional one: the one with the lowest dimension will be suppressed by the common scale $\Lambda$, while the others will be suppressed by powers of $\frac{H^3}{\Lambda^3} \cdot \sqrt{\eps} \frac{\Mpl}{H}$. The same
  conclusion applies to the minimal Galilean operators of cubic and quartic order, where there is a relation between quadratic and cubic terms in $\pi$. Therefore these operators have a negligible contribution to the three-point functions in our setup.

Since we are interested in quadratic and cubic operators and we do not have linear terms, we need to consider only terms with two traces on $\psi$ and terms with three traces:
\be
([\Psi \Psi \ldots \Psi] - c_1)([\Psi \Psi \ldots \Psi] - c_2) \qquad  {\rm and} \qquad ([\Psi \Psi \ldots \Psi] - c_3)([\Psi \Psi \ldots \Psi] - c_4)([\Psi \Psi \ldots \Psi] - c_5)
\ee
where the constants $c_i$ are chosen to subtract from each trace its background value. Operators that contain a $\Box \pi$ are proportional to the linear equation of motion and can be neglected since they can be removed by a field redefinition, as we discussed above. Of course this is valid only at leading order in slow-roll, but still tells us that operators with a $\Box \pi$ are subdominant with respect to the others.

The only cubic operator without $\Box \pi$, coming from a term with three traces is of course 
\be
(\nabla^2\pi)^3 \qquad \nabla^2 \equiv \delta^{ij} \nabla_i\nabla_j\pi \;.
\ee
Terms with two traces generate terms with 2 and 3 $\pi$'s and it is important to understand whether the latter are independent or not because this controls the scale that suppresses them, as we discussed in the previous paragraphs. If we restrict again to cubic operators without $\Box\pi$ the additional terms are  
\be
\nabla^2\pi (\nabla_i\nabla_j\pi)^2 \quad \nabla^2\pi (\nabla_i\nabla_\mu\pi)^2 \quad \nabla^2\pi (\nabla_\mu\nabla_\nu\pi)^2 \;.
\ee
It is straightforward to see that there is enough freedom to make all of them independent from the 2 $\pi$'s operators\footnote{This independence is not general: not all the cubic operators $(\partial^2\pi)^3$ are independent from the quadratic ones. It is straightforward, however, to check that the ones which are dependent contain a $\Box\pi$ and are thus neglected here. Another example is given by the two minimal Galilean operators that we are considering: also these are thus neglected in the calculation of the three-point function.}.

So far we have implicitly assumed that we can concentrate on the $\pi$ action, without taking into account its mixing with gravity. To check that this mixing is indeed negligible, one should follow \cite{Maldacena:2002vr}: in spatially flat gauge one solves the ADM constraints and plug the solution for the ADM variable $N$ and $N_i$ back into the action. To derive the cubic action, one needs to solve for $N$ and $N_i$ at first order only, as the second order result would multiply in the action the first order constraint equations, which vanish \cite{Maldacena:2002vr}. If the quadratic action is simply given by the standard kinetic term, then the mixing with gravity at energies of order $H$ is suppressed by the slow-roll parameters, so that the mixing with gravity just gives corrections to the 3-point function of the order of the slow-roll parameters.  In our case we also have additional quadratic operators -- see eq.~\eqref{eq:duckscan} -- of the schematic form $(\partial^2\pi_c)^2/\Lambda^2$. However, neglecting the mixing with gravity,  these terms are suppressed with respect to the canonical kinetic term by $H^2/\Lambda^2$, so that we expect they also give a suppressed contribution to the constraint equations. Indeed it is straigthforward to check that their effect is suppressed both by slow-roll and by $H^2/\Lambda^2$. Therefore we can safely concentrate on the $\pi$ action, without bothering about the mixing with gravity, to calculate the observable predictions. 

At third order in perturbations, we have the following four independent operators:
\begin{align}
S =& \int \rmd^4 x a^3 \bigg[ \Mpl^2 \dot H \de_\mu \pi \de^\mu \pi
+ \tilde{M}_1 (g^{ij} \nab_i \nab_j \pi)^3
+ \tilde{M}_2 (g^{ij} \nab_i \nab_j \pi) (\nab_i \nab_j \pi)^2 \notag \\
&+ \tilde{M}_3 (g^{ij} \nab_i \nab_j \pi) (\nab_i \nab_\mu \pi)^2
+ \tilde{M}_4 (g^{ij} \nab_i \nab_j \pi) (\nab_\mu \nab_\nu \pi)^2 \bigg],
\end{align}
in which we need to expand the covariant derivatives:
\begin{gather}
g^{ij} \nab_i \nab_j \pi = \frac{\de_i^2 \pi}{a^2} - \frac{1}{a^2} \del^{ij} \Gam^0_{ij} \dot{\pi}
= \frac{\de_i^2 \pi}{a^2} - 3 H \dot{\pi}\,, \\
(\nab_i \nab_j \pi)^2 = \frac{1}{a^4} \del^{ik} \del^{jl} (\de_i \de_j \pi - H a^2 \del_{ij} \dot{\pi}) (\de_k \de_l \pi - H a^2 \del_{kl} \dot{\pi})
= \frac{(\de_i \de_j \pi)^2}{a^4} - 2 H \dot{\pi} \frac{\de_i^2 \pi}{a^2} + 3 H^2 \dot{\pi}^2 \,,\\
(\nab_i \nab_0 \pi)^2 = - \frac{1}{a^2} (\de_i \dot{\pi} - H \de_i \pi)^2
\end{gather}
As discussed in the previous subsection, in order to calculate the non-Gaussian correlation functions at lowest order in slow-roll one can substitute the linear equation of motion $\ddot{\pi} + 3 H \dot{\pi} = \nabla^2 \pi / a^2$ in the action.
Substituting the Laplacian in favour of the time derivatives (in order to recover the usual shapes involving the operators $\dot{\pi}^3$ and $\dot \pi (\de_i \pi)^2/a^2$) yields
\begin{gather}
g^{ij} \nab_i \nab_j \pi = \ddot{\pi} \; , \qquad
(\nab_i \nab_j \pi)^2 = \frac{(\de_i \de_j \pi)^2}{a^4} 
- 2 H \dot{\pi} \ddot{\pi} - 3 H^2 \dot{\pi}^2 \; ,
\end{gather}
which results in an action with only three, and not four, independent operators. Doing a further integration by parts, $a^3 \ddot{\pi} \dot\pi^2 = \frac{1}{3} a^3\de_t \dot{\pi}^3 \rightarrow - a^3 H \dot{\pi}^3$, we finally get
\begin{align}
S =& \int \rmd^4 x \,a^3 \bigg[ -\Mpl^2  \dot H \left( \dot{
\pi}^2 - \frac{(\de_i \pi)^2}{a^2} \right)
  + M_1 \ddot{\pi}^3
  + M_2 \ddot\pi \frac{(\de_i \dot{\pi} - H \de_i \pi)^2}{a^2}
 \notag \\
& + M_3 \left(\ddot{\pi}  \frac{ (\de_i \de_j \pi)^2}{a^4} - 2 H
  \dot{\pi} \ddot{\pi}^2 + 3 H^3 \dot{\pi}^3 \right)
\bigg] \; .
\label{eq:Sint}
\end{align}
This will be the starting point for the calculation of the three-point
function.



\section{\label{sec:shapes}Shapes of non-Gaussianity}


Using the cubic action we can calculate the tree-level contribution to the three-point correlation function for the $\pi$ field. 
With the prescriptions of the \emph{in-in} formalism, this is given by the following formula~\cite{Maldacena:2002vr, Weinberg:2005vy}:
\beq
\avg{\pi^3(t_0)} = i \int_{-\infty}^{t_0} \rmd t \,a^3
\avg{\left[\pi^3(t_0), \int \rmd^3 x L_{\rm int}(t,\x) \right]}
\eeq
where $t_0$ is some late time when the physical modes of interest are well outside the Hubble radius, 
and $L_{\rm int}$ is the cubic Lagrangian to be read from Eq.~\eqref{eq:Sint}.

Non-Gaussianity is usually computed in terms of the correlation functions of the Bardeen potential $\Phi$. 
It is customary to write the correlation functions for $\Phi$ isolating the Dirac's delta that enforces the conservation of the three-momentum (implied by the traslational invariance). In the limit of exact scale invariance, the two- and three-point functions can be parametrized as
\begin{equation}
  \avg{\Phi_{\k_1} \Phi_{\k_2}} = (2 \pi)^3 \delta^{(3)}(\k_1+\k_2)
  \frac{\Delta_\Phi}{k^3}
\end{equation}
and
\begin{equation}
  \avg{\Phi_{\k_1} \Phi_{\k_2} \Phi_{\k_3}} = (2 \pi)^3
  \delta^{(3)}(\k_1+\k_2+\k_3) F(k_1, k_2, k_3) \; , 
\end{equation}
where $F$ (called the bispectrum of $\Phi$) is a homogeneous function of degree -6 which, because of rotational invariance, depends only on the length of the three-momenta. Therefore, in the same limit $k_1^6 F(k_1, k_2, k_3)$ becomes just a function of the ratios $r_2 = k_2/k_1$ and $r_3 = k_3/k_1$. As argued in \cite{Babich:2004gb}, the most meaningful quantity to plot is
\beq
r_2^2 r_3^2 F(1, r_2, r_3) \; , 
\eeq
which is also commonly called the ``shape'' of the non-Gaussianity.

During matter dominance $\Phi$ is related to the curvature perturbation on comoving slices $\zeta$ by the simple relation $\Phi=\frac{3}{5}\zeta$. The curvature perturbation is constant outside the Hubble radius, regardless of the matter content of the Universe (and in particular it remains constant during reheating). At leading order in slow roll $\zeta$ is related to the $\pi$ field by $\zeta = - H \pi$.
Shifting to conformal time ($\tau=-1/aH$), the bispectrum $F$ reads thus
\beq
\label{eq:bispectrum}
F(k_1, k_2, k_3) =
- \frac{i}{H} \bigg(\frac{3}{5}\bigg)^3  
\pi^*(0,k_1)\pi^*(0,k_2)\pi^*(0,k_3) \int_{-\infty}^0 
\frac{\rmd \tau}{\tau^4} \,
L_\mathrm{int}(\tau,k_1,k_2,k_3) + \mathrm{c.c.} + \mathrm{perms.}\; ,
\eeq
where $\pi(\tau, k)$ is the classical Fourier mode in de Sitter space (see Appendix \ref{app:shapes}) and the momenta are restricted to configurations forming a triangle in momentum space.

Doing this for each of the three independent cubic interactions in \eqref{eq:Sint} leads to three different bispectra, which are proportional to $M_1$, $M_2$ and $M_3$ respectively. As detailed in App.~\ref{app:shapes}, their form is (setting $k_t\equiv k_1+k_2+k_3$)
{\allowdisplaybreaks
 \begin{align}
  F_{M_1}(k_1, k_2, k_3) =
  &- \frac{20}{3} \, \frac{M_1H}{\eps \Mpl^2} \, 6\Delta_\Phi^2 
  \frac{k_t^3 -6k_t(k_1k_2+k_2k_3+k_1k_3)
  +15 k_1k_2k_3}{k_1 k_2 k_3 k_t^6}\;,
  \label{eq:FM1} \\
  F_{M_2}(k_1, k_2, k_3)
  = &\frac{5}{6} \, \frac{M_2H}{\eps \Mpl^2} \, \Delta_\Phi^2 
  \frac{k_1^2-k_2^2-k_3^2}{k_1 k_2^3 k_3^3 k_t}
  \bigg[2 + \frac{2(k_2 + k_3) - k_1}{k_t}
  \notag \\
  &-2\frac{k_1(k_2 + k_3) + 2 k_2^2 - 2 k_2 k_3 + 2 k_3^2 }{k_t^2}
  + 6\frac{k_1(k_3^2 - k_2k_3 + k_2^2) - 2 k_2^2k_3 - 2 k_2k_3^2}{k_t^3}
   \notag \\
  &+ 24\frac{k_1 k_2 k_3(k_2 + k_3) +2 k_2^2k_3^2}{k_t^4}
  - 120\frac{k_1 k_2^2 k_3^2}{k_t^5} 
  \bigg] + \mathrm{cyclic~permutations}\;,
  \label{eq:FM2} \\ 
  F_{M_3}(k_1, k_2, k_3)
  = &\frac{5}{3} \,\frac{M_3H}{\eps \Mpl^2} \,
  \frac{\Delta_\Phi^2}{k_1 k_2 k_3 k_t^3}
  \bigg[10-24\frac{k_2+k_3}{k_t}+48\frac{k_2k_3}{k_t^2} \notag \\
  &+\bigg(\frac{k_1^2-k_2^2-k_3^2}{k_2k_3}\bigg)^{\!2}
  \!\bigg(1 + 3\frac{k_2+k_3-k_1/2}{k_t} - 6\frac{k_1(k_2+k_3) - 2k_2k_3}{k_t^2}
  -30\frac{k_1k_2k_3}{k_t^3}\bigg)
  \bigg] \notag \\
  &+ \mathrm{cyclic~permutations}\;.
  \label{eq:FM3} 
\end{align}}

The simplest shape of non-Gaussianity is the so-called local shape, which peaks in the squeezed limit ($r_2\simeq1$ and $r_3\simeq 0$, \emph{i.~e.~}when a mode is much larger than the other two). In single field inflation non-Gaussianity of this kind is proportional to the tilt of the power spectrum, and a measurable signal can be generated only within multi-field scenarios \cite{Creminelli:2004yq}. Single field inflation can generate large non-Gaussianity when derivative interactions are present, whose shape is enhanced in other regions of the $r_1-r_2$ plane than the local limit. The shape generated in scenarios like DBI inflations, for instance, peaks in the equilateral limit ($r_2\simeq r_3 \simeq 1$) \cite{Babich:2004gb}. 
Shapes peaking in the ``flat'' configuration ($r_2\simeq r_3 \simeq 0.5$, corresponding to flattened isosceles triangles) where obtained in Ref.~\cite{Senatore:2009gt} as a difference of equilateral shapes, and directly from new higher derivative operators in Ref.~\cite{Bartolo:2010bj}. Notice however that Ref.~\cite{Bartolo:2010bj} did not identify any symmetry that allows to neglect lower dimensional operators and that operators containing higher time derivatives were not considered.

\begin{figure}[!htb]
\begin{center}
\begin{minipage}{.4\textwidth}
\includegraphics[width=\textwidth]{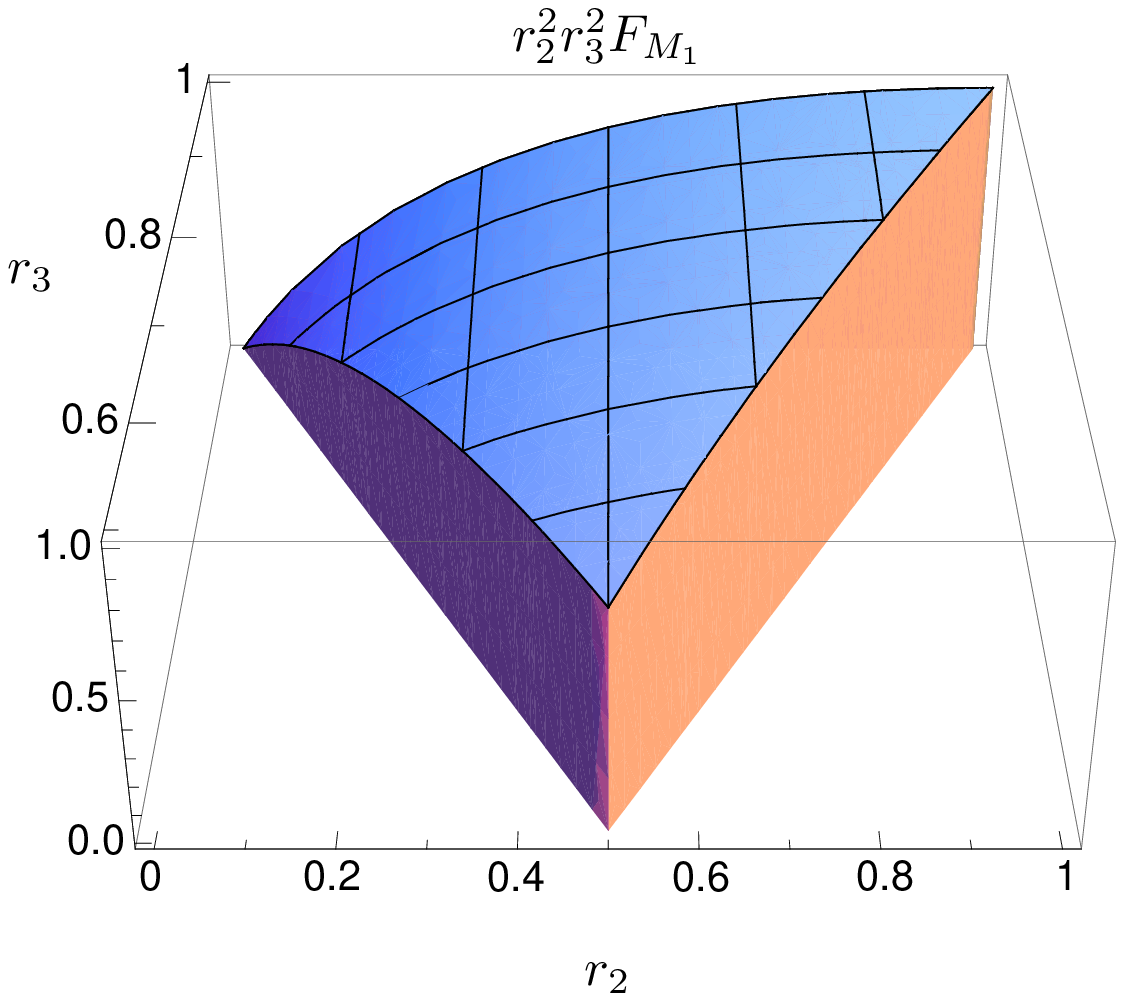}
\end{minipage}
\hspace{.1\textwidth}
\begin{minipage}{.4\textwidth}
\includegraphics[width=\textwidth]{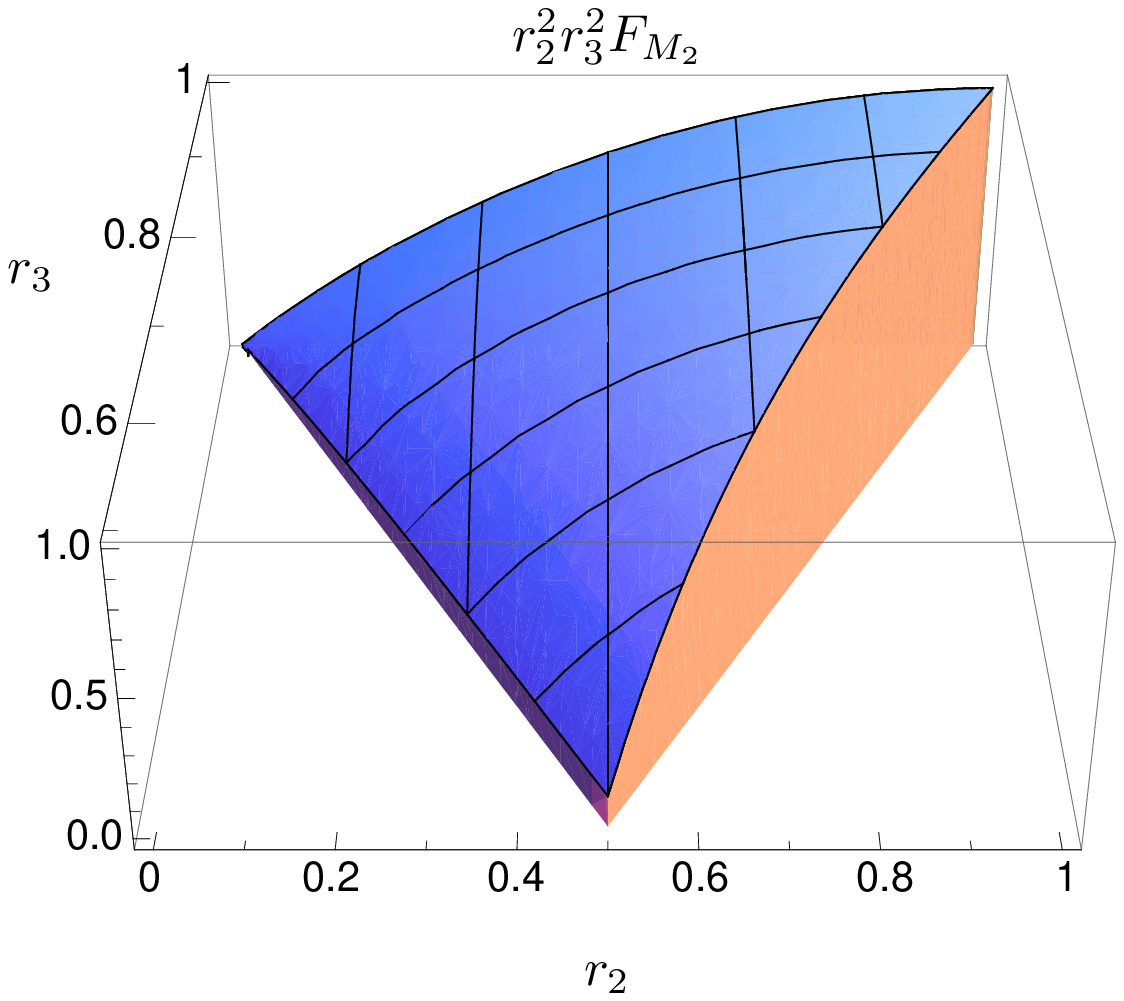}
\end{minipage}
\vspace{0.5cm}
\hspace{.1\textwidth}
\begin{minipage}{.4\textwidth}
\includegraphics[width=\textwidth]{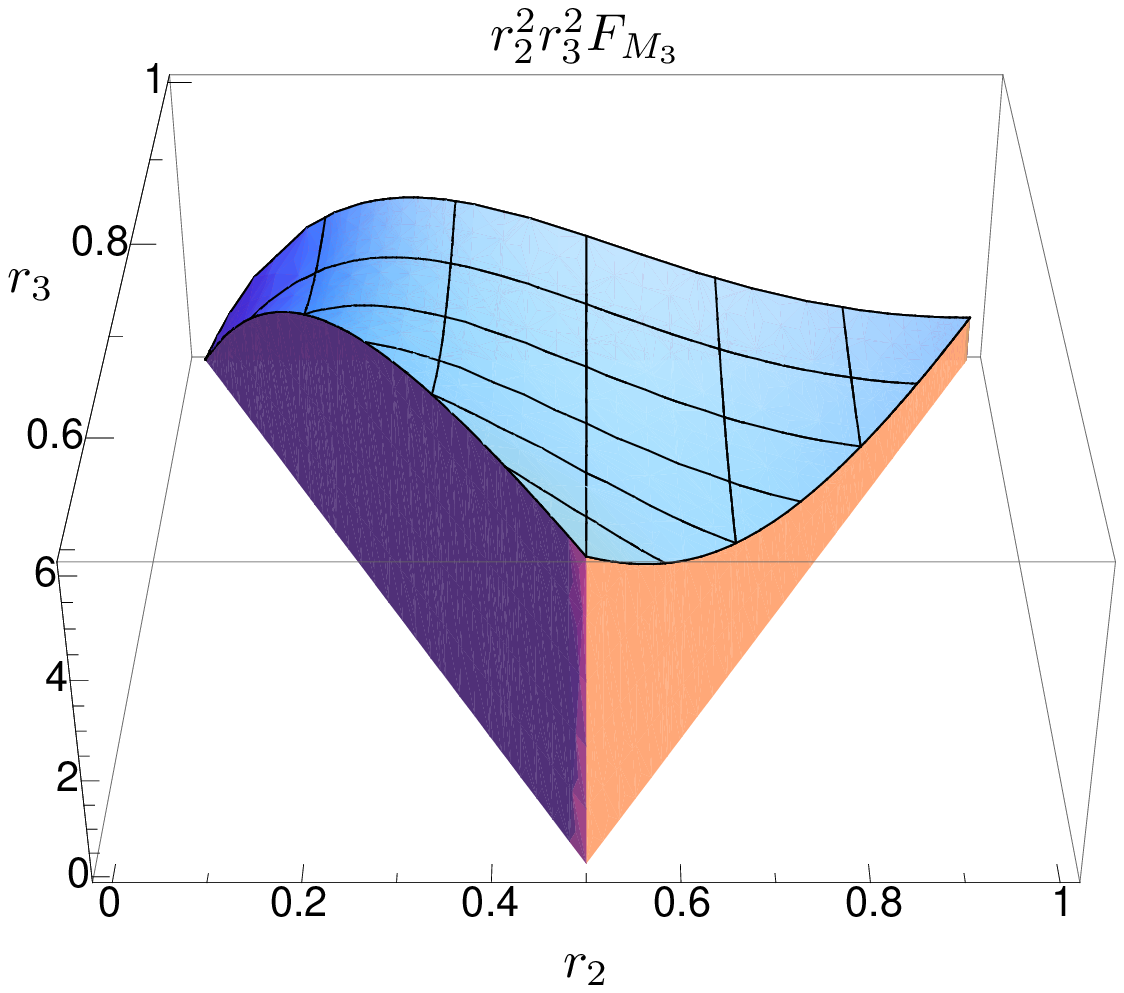}
\end{minipage}
\hspace{.1\textwidth}
\begin{minipage}{.4\textwidth}
\includegraphics[width=\textwidth]{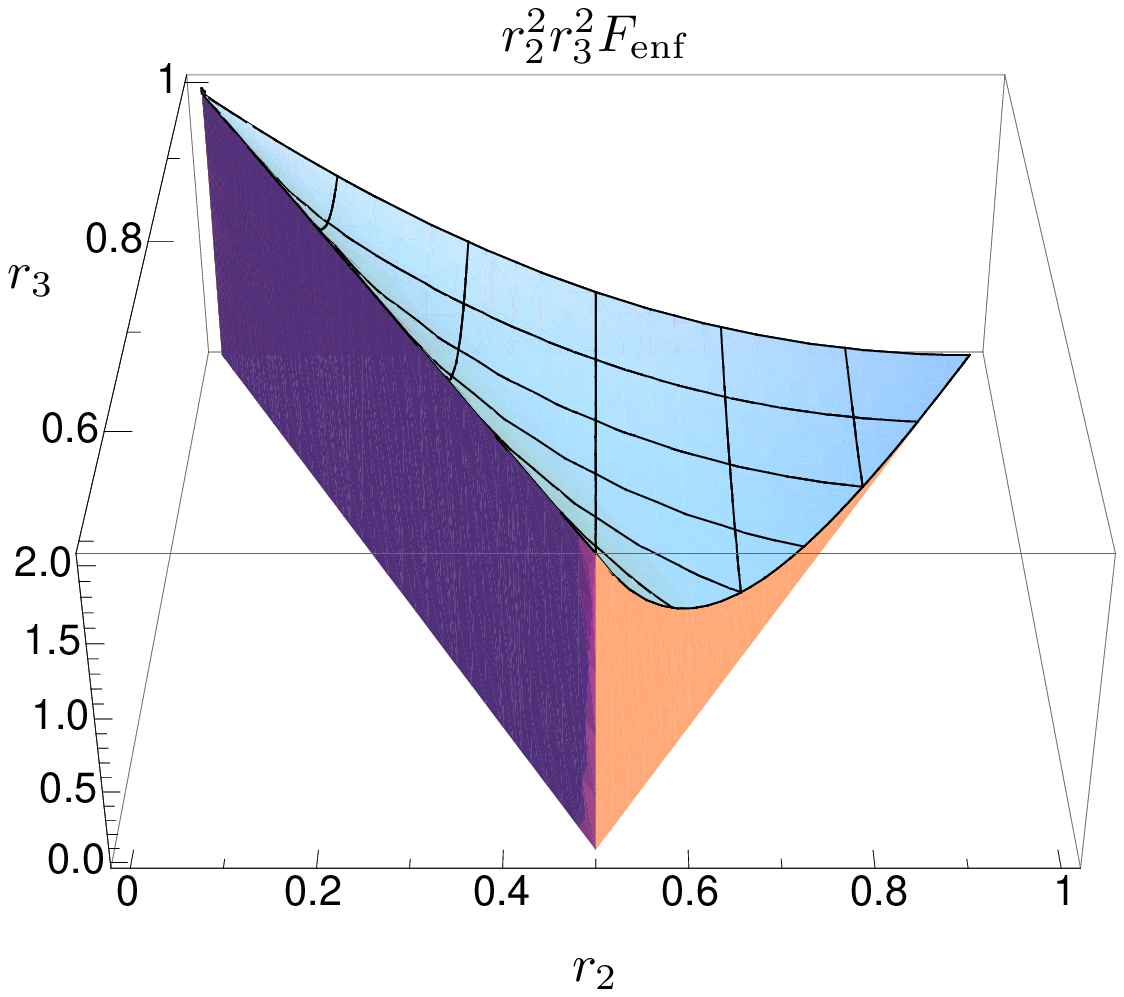}
\end{minipage}
\caption{\label{fig:shapes}\small Bispectrum shapes generated by the cubic
  operators in Eq.~\eqref{eq:Sint}. The ones proportional to $M_1$ and $M_2$ (upper panels)
  are very similar (with a cosine of nearly 1) to those generated by
  $\dot\pi^3$ and $\dot\pi(\de_i\pi)^2$ respectively, which where studied in
   \cite{Alishahiha:2004eh,Chen:2006nt,Cheung:2007st}. They have a cosine of 0.96 and 0.99 with the
  equilateral template. The third operator (lower left panel) is ``surfing'', i.e.~it is shaped
  like a wave with a maximum in the flat configuration, and has a large
  overlap with the enfolded template (lower right panel).
  The suppression in the equilateral regime is due to the presence of a 
  higher number of scalar products of gradients.}
\end{center}
\end{figure}

The shapes given by the three independent cubic operators in Eq.~\eqref{eq:Sint} are shown in Fig.~\ref{fig:shapes}.
The first operator $\ddot\pi^3$ has a clearly equilateral shape and is almost indistinguishable from the operator $\dot\pi^3$: indeed, in conformal time $\dot\pi_k\sim k^2\tau^2e^{i k \tau}$ is very similar to $\ddot\pi_k\sim k^2\tau^2(1+ik\tau/2) e^{i k \tau}$. For the same reason, $\ddot\pi(\de_i\dot\pi -H\de_i\pi)^2$ is very close $\dot\pi(\de_i\pi)^2$. Both operators have therefore equilateral shapes of the kind already considered in Refs.~\cite{Alishahiha:2004eh,Chen:2006nt,Cheung:2007st}.

On the other hand, an asymmetric operator with more scalar product of space gradients like $\ddot\pi(\partial_i\partial_j\pi)^2/a^4$ will contain terms like $(\k_2\cdot\k_3)^2=\frac{1}{4}(k_1^2-k_2^2-k_3^2)^2$. Moreover, each power of $1/a\sim\tau$ yields $1/(k_1+k_2+k_3)$ after the integral over conformal time; scalar products of gradients will therefore result in terms like $[(k_1^2-k_2^2-k_3^2)/2(k_1+k_2+k_3)^2]^n$, which tend to be suppressed in the equilateral limit. 
The precise scaling depends on the details of the symmetrization, but the general argument qualitatively holds. As a result, the shape generated by the operator proportional to $M_3$ looks like a wave reaching its maximum in the flat configuration.

In order to give a quantitative measure of the difference of the shapes generated by different interactions, it is very useful to define a scalar product between bispectra as follows~\cite{Babich:2004gb}:
\beq
F_{(1)} \cdot F_{(2)} = \sum_{\rm triangles} \frac{F_{(1)}(k_1, k_2, k_3) F_{(2)}(k_1, k_2, k_3)}{P(k_1) P(k_2) P(k_3)}
\label{eq:scalar}
\eeq
where the sum is restricted to the vectors $k_i$ that form a triangle in momentum space\footnote{For CMB applications it is useful to introduce a 2d scalar product that quantifies the similarity of two shapes in a given CMB experiment \cite{Babich:2004gb}. Here we stick to the 3d definition.}, from which it is natural to define the cosine
\beq
\cos(F_{(1)}, F_{(2)}) = \frac{F_{(1)} \cdot F_{(2)}}{(F_{(1)} \cdot F_{(1)})^{1/2} (F_{(2)} \cdot F_{(2)})^{1/2}} \; .
\eeq

If the cosine between two shapes is large, they have a significant overlap and it is possible to use the same estimator to analyze CMB data and constrain the amplitude $\fnl$ of each shape. On the other hand, if the cosine is small this is no longer efficient and a different estimator is needed in order to set optimal constraints on $\fnl$.

Moreover, a crucial numerical boost in CMB data analysis is gained by using factorizable shapes, \emph{i.e.}~shapes which can be written as sums of monomials of $k_1$, $k_2$ and $k_3$ \cite{Creminelli:2005hu, Wang:1999vf}. 
A good estimator to constrain the amplitude of a given model is then provided by a factorizable shape having a large cosine with the bispectrum generated by the model itself. Factorizable shapes which resemble model predictions are often called templates. 
Among the most popular templates are the local and equilateral ones \cite{Babich:2004gb}, for which CMB constraints are usually cited (see for example \cite{Komatsu:2010fb}). Linear combinations of equilateral operators can however yield a significantly different shape for a quite wide range of coefficients, leading to the definition of an ``orthogonal'' template \cite{Senatore:2009gt}. Another template found in the literature is the one for the so-called ``enfolded'' shape, originally introduced in Ref.~\cite{Meerburg:2009ys}. This template is a linear combination of the orthogonal and equilateral templates, as pointed out in Ref.~\cite{Senatore:2009gt}:
\begin{align}
F^\mathrm{enf}(k_1, k_2, k_3) 
&= \frac{1}{2}\left[F^\mathrm{equil}(k_1, k_2, k_3)- F^\mathrm{orth}(k_1, k_2, k_3)\right] \notag \\
&= 6\fnl^\mathrm{enf}\Delta_\Phi^2\Bigg[\frac{1}{k_1^3 k_2^3} + \frac{1}{k_1^2 k_2^2 k_3^2} - \frac{1}{k_1 k_2^2 k_3^3} - \frac{1}{k_2 k_1^2 k_3^3} + \mathrm{cyclic\; permutations}\Bigg]\,.
\label{eq:enfoldedTemplate}
\end{align}

We have computed the cosines of our shapes with the local, equilateral, orthogonal and enfolded templates. Our results are summarized in Table \ref{tab:cosines}.
\begin{table}
 \begin{center}
 \begin{tabular}{|c|c|c|c|c|}
        \hline 
        Shape: & local & equilateral & orthogonal & enfolded\\
        \hline         \hline 
        M1 & 0.49 & 0.96 & -0.06 & 0.70\\
        \hline
	M2 & -0.43 & -0.99 & -0.16 & -0.53\\
        \hline
	M3 & 0.57 & 0.71 & -0.52 & 0.94\\
        \hline
  \end{tabular}
  \end{center}
  \caption{\label{tab:cosines} \small Cosines between different shapes of bispectra.}
  \end{table}
As seen in this table, a suitable template for the operators $\ddot\pi^3$ and $\ddot\pi(\de_i\dot\pi -H\de_i\pi)^2$ is the equilateral one. The remaining operator has a large overlap with the enfolded template\footnote{There exists a linear combination of the equilateral and orthogonal templates, $F \propto -0.5 F^{\mathrm{orth}} + 0.86 F^{\mathrm{equil}}$, which has an even larger cosine (0.98) with $F_{M_3}$. However, it is good enough for our purpose to stick to the enfolded template, which has a cosine of 0.94 and is already known in the literature.}. Comparing with the templates allows us to compute the amplitude of non-Gaussianity of each type generated within our model, and hence constrain $M_1$, $M_2$ and $M_3$.

For nearly equilateral shapes, the amplitude of the non-Gaussianity is defined as $f_{\mathrm{NL}}^{\mathrm{eq}}\equiv k^6 F(k,k,k)/6\Delta_\Phi^2$ \cite{Creminelli:2005hu}. The amplitude induced by the two shapes $F_{M_1}$ and $F_{M_2}$ is thus
\begin{equation}
  f_{\mathrm{NL}}^{\mathrm{eq},1} = \frac{80}{729} \, \frac{M_1H}{\eps \Mpl^2}
  \,, \qquad
  f_{\mathrm{NL}}^{\mathrm{eq},2} = - \frac{865}{2916} \, \frac{M_2H}{\eps \Mpl^2}
  \,;
\end{equation}
using the most up-to date values $f_{\mathrm{NL}}^{\mathrm{eq}}=26\pm 140$ (68\% CL) from Ref.~\cite{Komatsu:2010fb} one gets 
\begin{equation}
  \frac{M_1H}{\eps \Mpl^2} = 240 \pm 1280\;, \qquad
  \frac{M_2H}{\eps \Mpl^2} = -80 \pm 470\;.
\end{equation}

For the shape $F_{M_3}$ the amplitude must be defined starting from the enfolded template,
\begin{equation}
  f_{\mathrm{NL}}^{\mathrm{enf},3} 
  \equiv \frac{k^6 F_{M_3}(k,k/2,k/2)}{96\Delta_\Phi^2} 
  = \frac{35}{256} \, \frac{M_3H}{\eps \Mpl^2} \, .
\end{equation}
Since this template is a linear combination of the equilateral and orthogonal ones (which are slightly correlated), the errors on the value of $\fnl^{\mathrm{enf}}$ can be inferred from the errors on $\fnl^{\mathrm{eq}}$ and $\fnl^{\mathrm{orth}}$ given in Ref.~\cite{Komatsu:2010fb}, assuming that the correlation coefficient remained the same as in Ref.~\cite{Senatore:2009gt}.  The limit obtained in this way is
\begin{equation}
f_\mathrm{NL}^\mathrm{enf} = 114 \pm 72 \quad (68\% \;\mathrm{CL}) \;,
\label{eq:limits}
\end{equation}
leading to 
\begin{equation}
  \frac{M_3H}{\eps \Mpl^2} = 830 \pm 530\;.
\end{equation}

A drawback of the enfolded template (as well as of the orthogonal one) is that it tends to a constant in the squeezed limit, while in the approximation of exact scale invariance any single-field three-point function vanishes in this regime \cite{Creminelli:2004yq}. 
This means that, while acceptable for CMB data analysis, such templates cannot give satisfactory results in contexts where all the signal is concentrated in the squeezed configurations as, for instance, with halo bias observations \cite{Catelan:1997qw,Dalal:2007cu,Verde:2009hy}. In order to improve the above constraint by combining it with LSS observations is therefore important to devise a different template, that peaks in the flattened limit but preserves the correct behavior when $k_1\ll k_2\sim k_3$. An example of such a template is given in App.~\ref{app:template}.

In our theory there are 3 independent parameters which allow us to choose arbitrary linear combinations of our operators. Fixing the amplitude of non-Gaussianity leaves us with a 2-parameter family ($M_2/M_1$ and $M_3/M_1$) of different possible shapes. It is interesting to see whether one can generate a suitable linear combination that is nearly orthogonal to all of the existing templates, following the same reasoning that lead to the definition of the orthogonal template in Ref.~\cite{Senatore:2009gt}. If this were the case, one should use a dedicated template for data analysis.

\begin{figure}
 \begin{center}
\begin{minipage}{.5\textwidth}
 \includegraphics[width=.9\textwidth]{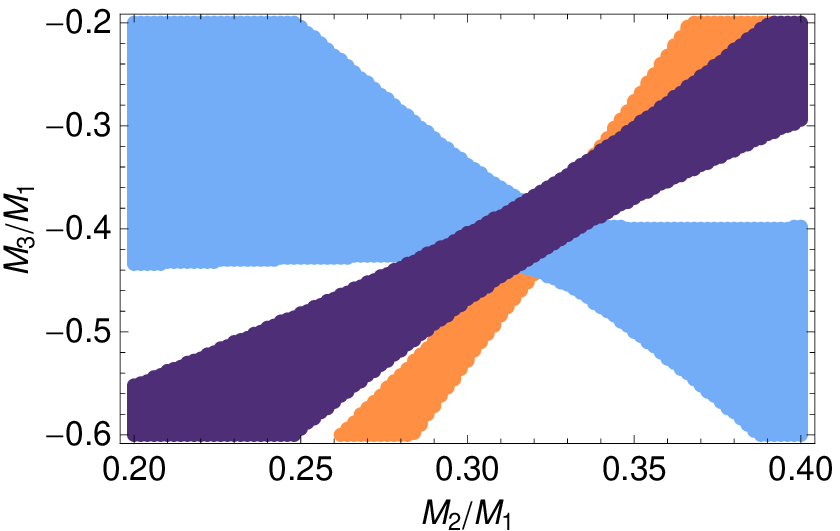}
\end{minipage}
\begin{minipage}{.45\textwidth}
 \hspace{.05\textwidth}
 \includegraphics[width=.9\textwidth]{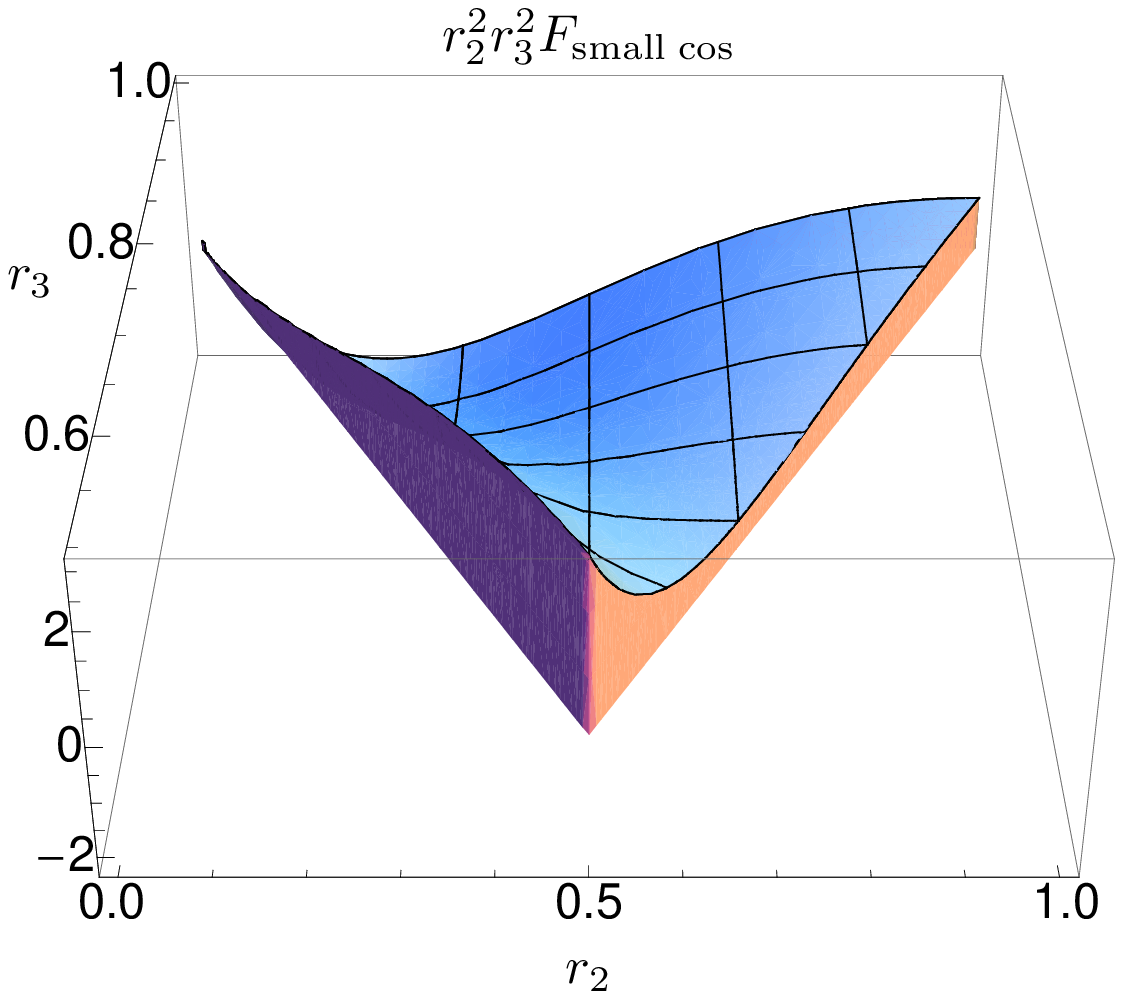}
\end{minipage}
 \end{center}
 \caption{\label{fig:cosines} \small \emph{Left:} regions of the parameter
 space $M_2/M_1$ and $M_3/M_1$ for which the cosine of the resulting
 shape with the local, equilateral and orthogonal template is smaller
 than 0.2. \emph{Right:} Non-Gaussian shape obtained with $M_2=0.32
 M_1$ and $M_3=-0.42 M_1$. This shape is nearly orthogonal to all known
 templates (with cosines of -0.15, 0.03, 0.06 and -0.03 with the local,
 equilateral, orthogonal and enfolded template), and would require a
 dedicated template for data analysis.}
\end{figure}

If such a linear combination exists, the vector $(1,M_2/M_1,M_3/M_1)$ must be an approximate null eigenvector of the $3\times3$ matrix of the scalar products of the three shapes (with the $M_i$'s set to 1) with the local, equilateral and orthogonal templates (the enfolded is already a linear combination of two of them). Equivalently, the coefficients $(M_i/M_1)\|F_{M_i}\|$ must form an approximate null eigenvector of the $3\times3$ matrix of the cosines, which are just normalized scalar products. Since the determinant of the matrix is 0.0057 (much smaller than most of the entries), such approximate eigenvector can be found and this allows to solve for $M_2/M_1$ and $M_3/M_1$. For instance, a linear combination with $M_2=0.32 M_1$ and $M_3=-0.42 M_1$ is a good candidate, yielding cosines of -0.15, 0.03 and 0.06 with the local, equilateral and orthogonal template (and -0.03 with the enfolded template).
The values of $M_2/M_1$ and $M_3/M_1$ yielding small (i.e.~$<0.2$) cosines with each template are shown in Fig.~\ref{fig:cosines}, where the shape of the combination nearly orthogonal to all templates is also given.

\section{\label{sec:NG4}Four-point function}


Let us see what are the implications of our setup for the four-point function\footnote{The contribution to the four-point function of operators with higher spatial derivatives was considered in \cite{Bartolo:2010di}.}. The action, besides the quadratic and cubic operators we considered, will contain quartic operators
\be
(\partial\pi_c)^2 + \frac{1}{\Lambda^2}(\partial^2\pi_c)^2 + \frac{1}{\Lambda^5}(\partial^2\pi_c)^3 +  \frac{1}{\Lambda^8}(\partial^2\pi_c)^4  + \ldots
\ee
The deviation from Gaussianity induced by the quartic terms can be quantified comparing these with the kinetic term at scales of order $H$
\be
\label{eq:NG4}
{\rm NG}_4 \equiv \frac{\langle \zeta^4\rangle}{\langle \zeta^2\rangle^2} \simeq \left.\frac{{\cal L}_4}{{\cal L}_2} \right |_{E \sim H} \simeq \left(\frac{H}{\Lambda}\right)^8 \;.
\ee 
The quartic non-Gaussianity comes from operators of dimension 12, and it is therefore suppressed by $(H/\Lambda)^8$. Notice that in terms of the parameter $\tau_{\rm NL}$ used in the literature, we have ${\rm NG}_4 \simeq \tau_{\rm NL} \Delta_{\zeta }$, where $\Delta_\zeta \simeq 2 \times 10^{-9}$ gives the normalization of the scalar power spectrum. As we saw, for the three-point function we have analogously
\be
\label{eq:NG3}
{\rm NG}_3 \equiv \frac{\langle \zeta^3\rangle}{\langle \zeta^2\rangle^{3/2}} \simeq \left.\frac{{\cal L}_3}{{\cal L}_2} \right |_{E \sim H} \simeq \left(\frac{H}{\Lambda}\right)^5 \;;
\ee 
indeed cubic operators are of dimension 9 and ${\rm NG}_3 \simeq f_{\rm NL}  \Delta_{\zeta}^{1/2}$. Without additional ingredients, we see from eq.s \eqref{eq:NG4} and \eqref{eq:NG3} that the quartic non-Gaussianity is always smaller than the cubic one
\be
{\rm NG}_4 \sim {\rm NG}_3^{8/5} \;,
\ee
so that the three-point function is much easier to detect. Notice, however, that the scaling of the quartic and cubic non-Gaussianities is different from the one one has in models with a low speed of sound, where the three-point function is given by operators of dimension 6 (schematically $(\partial\pi_c)^3/\Lambda^2$), while the four-point function is given by operators of dimension 8 (of the form $(\partial\pi_c)^4/\Lambda^4$), so that
\be
{\rm NG}_3 \simeq  \left(\frac{H}{\Lambda}\right)^2 \qquad  {\rm NG}_4 \simeq  \left(\frac{H}{\Lambda}\right)^4  \quad\Rightarrow\quad {\rm NG}_4 \sim {\rm NG}_3^{2} \;. 
\ee
This implies that, at the same level of cubic non-Gaussianity, our setup gives a larger four-point function signal, although still much smaller than the cubic one. This difference is somewhat important experimentally. If the three-point function is of the order of the present experimental limit,  $f_{\rm NL} \sim 100$, equivalent to ${\rm NG}_3 \sim 5 \times 10^{-3}$, then in models with a small speed of sound one has $ {\rm NG}_4 \sim 2 \times 10^{-5}$, i.e.~$\tau_{\rm NL} \sim 10^4$, so that the four-point function is not detectable by the Planck satellite, while it may be detectable in our setup, where  $ {\rm NG}_4 \sim 2 \times 10^{-4}$, i.e.~$\tau_{\rm NL} \sim 10^5$. Obviously these are just rough estimates as a proper forecast for the Planck capability on the 4-point function is still lacking.  

The three-point function is usually considered the leading signal of non-Gaussianity as it typically dominates the higher order correlations as we saw above. However, it has been pointed out in \cite{Senatore:2010jy} that it is possible to have technically natural theories in which the leading source of non-Gaussianity is the four-point function, as a consequence of an approximate $\pi \to - \pi$ symmetry. We can straightforwardly apply the same arguments in our context. Let us consider the case in which, besides the standard kinetic term, we only have operators that start quartic in $\pi$. Of course both the kinetic term and these quartic interactions respect the symmetry $\pi \to - \pi$.  However, as a consequence of the non-linear realization of the Lorentz symmetry, these operators will also contain terms of higher order in $\pi$, in particular quintic, that do not respect the symmetry\footnote{Actually the operator $(\Box\psi+3H)^4$ does not contain quintic or higher order terms, so that the parity symmetry would be exact. However this operator gives a four-point function which is four times suppressed by slow-roll as all the legs are proportional to the linear equation of motion.}. Obviously if these terms were unsuppressed the symmetry would be of no use. However, as we discussed above, operators with a given dimension that are related to lower dimensional ones by the Lorentz symmetry are not weighed by the cut-off scale $\Lambda$, but they are further suppressed. In particular the quintic operators will be of the form
\be
\xi \frac{(\partial^2 \pi_c)^5}{\Lambda^{11}} \qquad \xi \equiv  \Lambda^5 \frac{H^3}{\Lambda^3} \cdot \sqrt{\eps} \frac{\Mpl}{H} \ll 1 \;.
\ee
The breaking of the $\pi \to -\pi$ symmetry will thus be weighted by the small parameter $\xi$. In particular loops will induce cubic operators of the form
\be
\xi \frac{(\partial^2 \pi_c)^3}{\Lambda^{5}} \;.
\ee
These radiatively induced cubic interactions only give a completely negligible
\be
{\rm NG}_3 \simeq \xi \left(\frac{H}{\Lambda}\right)^5  = \left(\frac{H}{\Lambda}\right)^2 \frac{H}{\Mpl \sqrt{\eps}} \simeq \left(\frac{H}{\Lambda}\right)^2 \Delta_\zeta^{1/2} \quad\Rightarrow \quad f_{\rm NL} \simeq  \left(\frac{H}{\Lambda}\right)^2 \;.
\ee
We conclude that it is possible in our setup to have a large four-point function, of order $(H/\Lambda)^8$, with a negligible three-point function signal. The number of independent operators -- the quartic operators that are independent of the cubic ones -- is quite large, even if we restrict to terms without $\Box\pi$. Their study is beyond the scope of this paper.


\section{\label{sec:conclusions}Conclusions and future directions}


In this paper we explored a new class of single field inflationary models, endowed with an approximate Galilean symmetry. Their phenomenology is quite distinctive, with new shapes of the three-point function and potentially large four-point function. 

The study of this class of theories can be extended in many directions. As we discussed, there are regions of parameter space in which the shape of the three-point function is quite different from all the ones for which a data analysis has been performed, so that the constraints on these regions are very loose. It would be important to extend the analysis of CMB maps to these shapes. Even more general shapes can be obtained considering, in addition to the $(\partial^2\pi)^3$ operators, operators with less derivatives (Galilean invariant or not). It is technically natural to keep these additional terms small, since they are not radiatively induced by the operators already included in the Lagrangian. Therefore they can give a contribution to the three-point function comparable to that of terms with more derivatives. In this paper we have just performed a quite preliminary and qualitative study of the four-point function signal. Since in these models this signal is larger than in models with a reduced speed of sound (for similar values of the three-point function) and it can become very large imposing an approximate $Z_2$ symmetry, we believe it is worthwhile to characterize the four-point signal and compare it with data. 

From the point of view of model building, it is interesting to note that one may go a step further and study a theory in which interactions start with three or more derivatives per field. Although this cannot be dictated by any symmetry -- the standard kinetic term is {\em not} invariant under a constant shift of the second derivative -- terms with less derivatives in the interactions will not be generated. It may be interesting to study the phenomenology of these higher derivative models. Another possible direction of further study is the multi-field scenario: as in the case of models with reduced speed of sound \cite{Langlois:2008qf}, multi-field models may have interesting signatures and it may be useful to explore the effects of the Galilean symmetry in the effective theory of multi-field inflation \cite{Senatore:2010wk}.

\section*{Ackowledgements}
It is a pleasure to thank Andrei Gruzinov, Lam Hui, Mehrdad Mirbabayi, Alberto Nicolis and Matias Zaldarriaga for useful discussions.
P.~C.~thanks Leonardo Senatore and Matias Zaldarriaga for stressing the effective field theory approach to inflation. The work of G. D'A. is supported by a James Arthur fellowship.


\begin{appendix}
\section{\label{app:NR}Non-renormalization of Galilean operators}


In this Appendix we complete the analysis of the non-renormalization properties of Galilean interactions in the presence of gravity. 
In Section \ref{sec:action} we discussed what happens when the leading Galilean terms of the form $(\partial \psi)^2 (\partial^2 \psi)^{n}$ are zero.
In flat space these operators are not renormalized, but one may wonder whether in the presence of an external gravitational field generic terms of the form $(\partial \psi)^4 (\partial^2 \psi)^{n-2} R$ can be generated. This would be relevant in our construction, since Galilean terms with the minimum number of derivatives were introduced to get rid of terms linear in $\pi$. We show in the following that operators of the form $(\partial \psi)^4 (\partial^2 \psi)^{n-2} R$ cannot appear, because vertices generated by loop diagrams with $n$ fields $\psi$ must have at least $2n$ derivatives also on curved backgrounds.  

To make clear the logic of our argument let us review first why this is the case in flat space.
We consider 1PI graphs that involve the leading Galilean interactions. Each vertex has at least 2 internal lines, otherwise the graph is not 1PI. We want to show that we can always take the two lines with only one derivative to be the internal ones, so that the resulting graph has more derivatives than the minimal Galilean vertex. Following \cite{Endlich:2010zj}, if we pick two internal lines and call them $\phi$ the vertices can be schematically of the form  
\begin{gather}
(\partial^2 \psi)^{n-2} \partial \phi \, \partial \phi  \label{didi} \\ 
\partial \psi (\partial^2 \psi)^{n-3} \, \partial^2 \phi \, \partial \phi   \label{di2di} \\
\partial \psi \partial \psi (\partial^2 \psi)^{n-4} \, \partial^2 \phi \, \partial^2 \phi \,  \label{di2di2}
\end{gather} 
however it is easy to see that we can always rewrite the last two in the form of the first, namely with only one derivative acting on each $\phi$.
For (\ref{di2di}) it is immediate since  
\be\label{intbyparts}
\partial_{\mu}\partial_{\nu}\phi \, \partial_{\alpha} \phi = \partial_{( \mu} (\partial_{\nu )} \phi \, \partial_{\alpha} \phi ) - \frac{1}{2} \partial_{\alpha} (\partial_{\mu} \phi \, \partial_{\nu} \phi)
\ee
and integrating by parts we can move one derivative to the other lines. 
The interaction (\ref{di2di2}) can be also be put in the form (\ref{di2di}). If we write it as $A^{\mu\nu\rho\sigma} \partial_\mu \partial_\nu \phi \, \partial_\rho \partial_\sigma \phi$, we see that it gives a contribution to the equation of motion of the form $A^{\mu\nu\rho\sigma} \partial_\mu \partial_\nu  \partial_\rho \partial_\sigma \phi$ that must vanish because the e.o.m. obtained from minimal Galilean terms has 2 derivatives per field. This is possible only if the totally symmetric part of $A^{\mu\nu\rho\sigma}$ is zero.
However, since the structure $ \partial_\mu \partial_\nu \phi \, \partial_\rho \partial_\sigma \phi$ has less symmetries under permutation of indices, it doesn't imply that the vertex equally vanishes. In particular, $A$ can be antisymmetric in the exchange $[ \mu \rho ]$ and symmetric in all the others, thus giving a vanishing contribution to the e.o.m. but a non-zero vertex. In this case anyhow we can write
\be\label{trederiv}
\partial_{\nu}\partial_{[\mu}\phi \, \partial_{\rho]} \partial_{\sigma} \phi = \partial_{[ \mu} (\partial_{\nu} \phi \, \partial_{\rho]}\partial_{\sigma}  \phi ) 
\ee 
and integrating by parts we end up with a vertex of the form (\ref{di2di}) and then use again (\ref{intbyparts}) to complete the argument.

Now we want to see what changes if we include an external gravitational field. The starting point is the minimal covariantization of the leading Galilean interactions and we consider again 1PI graphs with $n$ external fields $\psi$ plus external graviton lines.
There are two possibilities: when gravitons are attached to internal lines the vertices are the same as in flat space and we can proceed as before in order to finish with one derivative on each of the two $\phi$. This will simply add $g$ or $R$ or $\nabla R$ etc. to a $(\partial^2 \psi)^n$ graph.
If instead the gravitons lines go in the vertices we can try to repeat the same steps we did before with the substitution of ordinary derivatives with covariant ones. 
The extra derivative in the structure $\nabla \nabla \phi \, \nabla \phi$ can be moved to the other lines using the identity (\ref{intbyparts}) that is valid also with $\nabla$ instead of $\partial$ since it involves at most two derivatives on the same scalar field and so there is no problem in commuting covariant derivatives. The difference arises when we consider the vertex $A^{\mu\nu\rho\sigma} \nabla_\mu \nabla_\nu \phi \, \nabla_\rho \nabla_\sigma \phi$. Because $A$ contains at most two covariant derivatives on $\psi$ and no $R$, its symmetries are the same as in flat space: the only component different from zero can be taken to be antisymmetric in $[ \mu \rho ]$. 
Now though eq. (\ref{trederiv}) gets an extra piece proportional to the Riemann tensor from the commutator of covariant derivatives:   
\be
\nabla_{\nu}\nabla_{[\mu}\phi \, \nabla_{\rho]} \nabla_{\sigma} \phi = \nabla_{[ \mu} (\nabla_{\nu} \phi \, \nabla_{\rho]}\nabla_{\sigma}  \phi ) + R^{\epsilon}_{\mu\rho\sigma} \partial_\epsilon \phi \, \partial_\nu \phi \, . 
\ee
The first term can be integrated by parts as before while from the second we see that it is possible to trade two external derivatives in the vertex for a factor of $R$. However, the minimal number of derivatives with $n$ fields $\psi$ is still $2n$ and because of that a term of the form $(\partial \psi)^4 (\partial^2 \psi)^{n-4} R$ cannot be generated.


\section{\label{app:shapes}Calculation of the bispectra}


At leading order in slow roll the field $\pi_k(\tau)$ and its derivatives, which must be inserted in \eqref{eq:bispectrum} in order to compute the bispectrum, read in conformal time 
\begin{equation}
  \pi_k(\tau) = i\frac{1-i k \tau}{2\sqrt{\eps k^3}\Mpl}
  e^{ik\tau} \;,\quad
  \dot\pi_k(\tau) = -i \frac{H k^2\tau^2}{2\sqrt{\eps k^3}\Mpl}
  e^{ik\tau} \;, \quad
  \ddot\pi_k(\tau) = i \frac{H^2 k^2\tau^2}{2\sqrt{\eps k^3}\Mpl}
  (2+ik\tau)e^{ik\tau} \;.
\end{equation}
Hence, from all the modes in \eqref{eq:bispectrum} there will be a factor 
\begin{equation}
  \frac{1}{2^6 \eps^3 k_1^3k_2^3k_3^3 \Mpl^6}e^{i(k_1+k_2+k_3)\tau}
\end{equation}
which is common to all interactions. Additional factors for the different fields in each interaction must be accounted for according to the following rules:
\begin{equation}
  \pi \longrightarrow 1 - ik_i\tau \;,\qquad
  \dot\pi \longrightarrow -H k_i^2\tau^2 \;,\qquad
  \ddot\pi  \longrightarrow H^2 k_i^2\tau^2(2+ik_i\tau) \;, \qquad
  \frac{\vec \de}{a}  \longrightarrow -iH\vec k_i\tau \;.
\end{equation}
Since we are looking at interactions with six derivatives, there will also be an overall factor of $H^6$. Recalling the definition $\Delta_\Phi=\frac{9}{25}H^2/(4 \eps \Mpl^2)$, the generic bispectrum will thus look like
\begin{equation}
  F_{M_i}(k_1, k_2, k_3) = -i 
  \frac{5 M_iH}{12 \eps \Mpl^2}
  \frac{\Delta_\Phi^2}{k_1^3k_2^3k_3^3} \int_{-\infty}^0 \mathrm{d}\tau \,
  P_{M_i}(k_i,\tau)e^{ik_t\tau} + \mathrm{c.c.} + \mathrm{permutations} \,,
\end{equation}
where the function $P(k_i,\tau)$ contains all the appropriate additional factors for the interaction as specified above, divided by $H^6\tau^4$ (because of the factor of $\tau^{-4}$ in \eqref{eq:bispectrum} and the factor of $H^6$ we extracted).

It is convenient to shift to the adimensional integration variable $y=k_t\tau$, and introduce a dummy parameter $\mu$ in the phase ($e^{iy}\rightarrow e^{i\mu y}$) to be set to one at the end of the computation. By doing this one can substitute $\tau\rightarrow -(i/k_t)\frac{\mathrm{d}}{\mathrm{d}\mu}$ in the function $P$, which can thus be pulled out from the integration: all is left to do is therefore the simple integral of a phase, $\int_{-\infty}^0e^{i\mu y} = -i/\mu$.
After all these steps the final result is
\begin{equation}
  F_{M_i}(k_1, k_2, k_3) =
  - \frac{5 M_iH}{6 \eps \Mpl^2} \frac{\Delta_\Phi^2}{k_1^3k_2^3k_3^3 k_t}
  \bigg[\hat P_{M_i} \bigg(k_i,-\frac{i}{k_t}\frac{\mathrm{d}}{\mathrm{d}\mu}\bigg) \frac{1}{\mu}\bigg]_{\mu=1}
  + \mathrm{permutations}\,,
\end{equation}
where the factor of 2 comes from the complex conjugate (since the result is real). In other words, the specific momentum dependence of each interaction is contained in the operator $\hat P$ that must be applied to $1/\mu$ before setting $\mu=1$. The operator associated to a cubic interaction (but the argument can be generalized to any interaction) with $n$ space gradients is given by the composition of

\noindent
i) 3 differential operators according to the following substitutions:
\begin{equation}
  \pi \longrightarrow 
  1- \frac{k_i}{k_t} \frac{\mathrm{d}}{\mathrm{d}\mu}
   \;,\qquad
  \dot\pi \longrightarrow 
  \bigg(\frac{k_i}{k_t}\bigg)^{\!2} \frac{\mathrm{d^2}}{\mathrm{d}\mu^2}
   \;,\qquad
  \ddot\pi  \longrightarrow 
  - \bigg(\frac{k_i}{k_t}\bigg)^{\!2} \bigg(2+\frac{k_i}{k_t}\frac{\mathrm{d}}{\mathrm{d}\mu}\bigg) \frac{\mathrm{d^2}}{\mathrm{d}\mu^2}\;;
\end{equation}

\noindent
ii) $n$ operators $- (\vec k_i/k_t) \frac{\mathrm{d}}{\mathrm{d}\mu}$, one for each spatial derivative;

\noindent
iii) one operator $[(1/k_t)\frac{\mathrm{d}}{\mathrm{d}\mu}]^{-4}$ to account for $\tau^{-4}$.

The operators generated by the interactions studied in this paper are
\begin{gather}
  -\frac{k_1^2k_2^2k_3^2}{k_t^2}
  \left(2+\frac{k_1}{k_t}\frac{\mathrm{d}}{\mathrm{d}\mu}\right)\!
  \left(2+\frac{k_2}{k_t}\frac{\mathrm{d}}{\mathrm{d}\mu}\right)\!
  \left(2+\frac{k_3}{k_t}\frac{\mathrm{d}}{\mathrm{d}\mu}\right)
  \frac{\mathrm{d}^2}{\mathrm{d}\mu^2}, \\
  - k_1^2(\vec k_2\cdot\vec k_3)
  \left(2+\frac{k_1}{k_t}\frac{\mathrm{d}}{\mathrm{d}\mu}\right)\!
  \bigg[1-\frac{k_2}{k_t}\frac{\mathrm{d}}{\mathrm{d}\mu}
  -\bigg(\frac{k_2}{k_t}\bigg)^{\!2}\frac{\mathrm{d}^2}{\mathrm{d}\mu^2}\bigg]\!
  \bigg[1-\frac{k_3}{k_t}\frac{\mathrm{d}}{\mathrm{d}\mu}
  -\bigg(\frac{k_3}{k_t}\bigg)^{\!2}\frac{\mathrm{d}^2}{\mathrm{d}\mu^2}\bigg]
\end{gather}
and
\begin{equation}
  -\frac{k_1^2k_2^2k_3^2}{k_t^2}
  \bigg[\frac{(\vec k_2\cdot\vec k_3)^2}{k_2^2 k_3^2}\!
  \left(2+\frac{k_1}{k_t}\frac{\mathrm{d}}{\mathrm{d}\mu}\right)\!
  \left(1-\frac{k_2}{k_t}\frac{\mathrm{d}}{\mathrm{d}\mu}\right)\!
  \left(1-\frac{k_3}{k_t}\frac{\mathrm{d}}{\mathrm{d}\mu}\right)\!
  + 2\left(2+\frac{k_2}{k_t}\frac{\mathrm{d}}{\mathrm{d}\mu}\right)\!
  \left(2+\frac{k_3}{k_t}\frac{\mathrm{d}}{\mathrm{d}\mu}\right) - 3 \bigg]
  \frac{\mathrm{d}^2}{\mathrm{d}\mu^2}
\end{equation}
for $M_1$, $M_2$ and $M_3$ respectively. All of them must be applied to $1/\mu$ and evaluated at $\mu=1$.

The calculation of the different bispectra is now straightforward given the relation
\begin{equation}
  \bigg[\frac{\mathrm{d}^n}{\mathrm{d}\mu^n}\frac{1}{\mu}\bigg]_{\mu=1}
  = (-1)^n n!
\end{equation}
and gives the results quoted in Eqs.~\eqref{eq:FM1}, \eqref{eq:FM2} and \eqref{eq:FM3}.


\section{\label{app:template}A template that vanishes in the squeezed limit}


For the analysis of CMB data the enfolded template, equation \eqref{eq:enfoldedTemplate}, is good enough for constraining the non-Gaussianity induced by the operator proportional to $M_3$ in \eqref{eq:Sint}. Since the enfolded template is a linear combination of the equilateral and orthogonal templates, its amplitude can be constrained using the recent limits on these shapes,  as discussed in equation \eqref{eq:limits}.

However, the enfolded template fails to capture some important qualitative behaviour of the physical shape as we will now explain. There are competitive constraints on the non-Gaussianity from Large Scale Structure observations \cite{Slosar:2008hx}. These arise from the fact that the non-Gaussian corrections to the halo bias $\Delta b$ have a characteristic scale dependence and one expects this scale dependence to be sensitive to the squeezed limit of the non-Gaussian shape. Both the orthogonal and enfolded templates go to a constant in the squeezed limit, $\lim_{r_2\rightarrow 0}r_2^2 r_3^2 F(1,r_2,r_3) = const.$, and one could thus expect some scale dependent effect in the halo bias induced by this non-Gaussianity. However, physical shapes generated by the single-field inflation models studied in the literature go to zero in the squeezed limit \cite{Creminelli:2004yq, Cheung:2007st, Senatore:2009gt}. Indeed this can be explicitly checked in \cite{Senatore:2009gt} for the model from which the orthogonal shape was inspired, and for the model under study in this paper. This means that one expects the scale dependence of $\Delta b$ to be negligible for these models, even if they have a large cosine with templates which do induce a scale dependence.

This motivates us to introduce a new factorizable template which apart from having a large cosine with the physical shape also behaves like the physical shape in the squeezed limit\footnote{A similar approach has been taken in Appendix B of \cite{Senatore:2009gt}. Notice that in the published version of the paper, the proposed template contains some typos, that have been corrected in the most recent version. We thank L.~Senatore for correspondence about this point.}. Let us warn the reader though that it is not clear whether the templates introduced in this Appendix are useful for computing other quantities which are interesting for LSS studies, as they might be sensitive to specific limits where the physical shape and the template differ.

\begin{figure}
\begin{center}
\begin{minipage}{.4\textwidth}
\includegraphics[width=\textwidth]{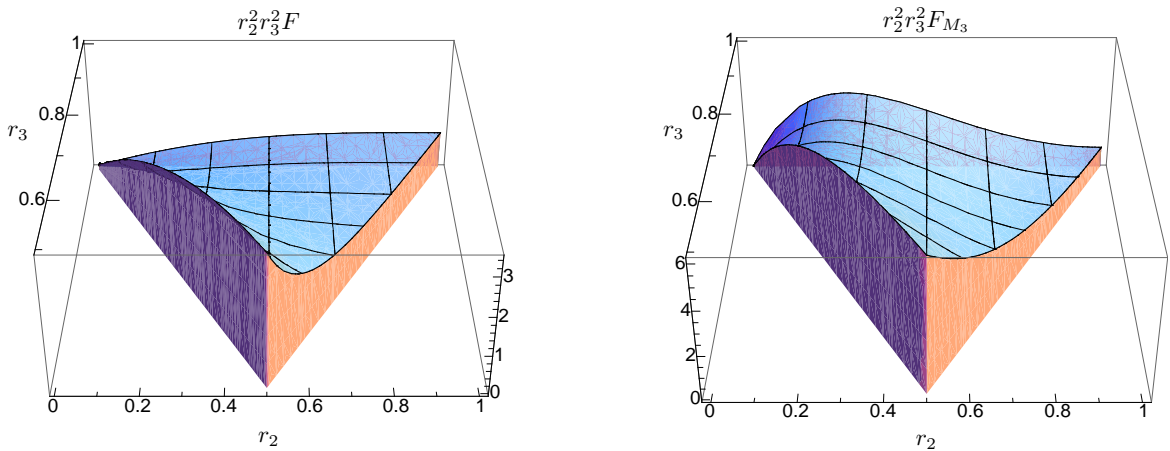}
\end{minipage}
\hspace{.1\textwidth}
\begin{minipage}{.4\textwidth}
\includegraphics[width=\textwidth]{M3.eps}
\end{minipage}
\caption{\label{app:fig:template}\small In the left panel this figure we show the template \eqref{app:eq:template} that goes to zero in the squeezed limit and has a large cosine with the physical bispectrum shape generated by the operator proportional to $M_3$, which is shown in the right panel for comparison.}
\end{center}
\end{figure}

The new template will contain also monomials that go like $k^{-4}$: this will give us additional freedom to impose that it behaves as $k_l^{-1}$ in the squeezed limit, $k_l \to 0$.
One way to build the template is to take a linear combination of the new monomials for which the leading divergencies $k_l^{-4}$ and $k_l^{-3}$ cancel. We do not have enough freedom to cancel also the $k_l^{-2}$ divergence, but we can at least require that the divergence is independent of the direction we approach the limit. This will allow a cancellation with the ``standard'' monomials. The only combination with such properties is given by
\begin{equation}
F_1(k_1,k_2,k_3) = \frac{16}{9 k_1 k_2 k_3^4} + \frac{k_1^2}{9k_2^4 k_3^4} - \frac{1}{k_1^2 k_3^4} - \frac{1}{k_2^2 k_3^4} + \mathrm{cyclic\;perms.}\,.
\end{equation}
Now we can  combine this with the usual monomials diverging as $k^{-3}$ and $k^{-2}$ 
\begin{align}
F_2(k_1,k_2,k_3) &= \frac{1}{k_1^3 k_2^3} - \frac{1}{k_1 k_2^2 k_3^3} - \frac{1}{k_2 k_1^2 k_3^3} + \mathrm{cyclic\;perms.}\,,\\
F_3(k_1,k_2,k_3) &= \frac{1}{k_1^2 k_2^2 k_3^2}\,.
\end{align}
The residual $k_l^{-2}$ divergence of $F_1$, which does not depend on the direction, can be cancelled by a proper addition of $F_3$. This gives a new template, besides the standard equilateral one, with the correct squeezed limit. We have thus a one-parameter family of templates going as $k_l^{-1}$ in the squeezed limit:
\begin{equation}
F = A \fnl\Delta_\Phi^2\left[\alpha F_1 + F_2 + 2(1 + \alpha) F_3\right]\,,
\label{app:eq:template}
\end{equation}
where $\alpha$ is a free coefficient which can be fixed by requiring the cosine with the physical shapes to be maximum, and $A$ is some normalization which can be fixed, for example, such that the template equals the local one in the equilateral limit\footnote{The total number of monomials is $7$, so that we have 6 free parameters and an overall normalization. Requiring a $k_l^{-1}$ divergence implies 4 conditions, leaving a 2 parameter set of templates with the correct squeezed behaviour. The construction given in the text corresponds to a particular one-parameter subset of templates, which turns out not to contain the monomial $k_1/(k_2^4 k_3^3) +$perms. In the latest version of \cite{Senatore:2009gt} a different one-parameter set of templates is given.}. The equilateral template corresponds to $\alpha = 0$. In table \ref{tab:alpha} we show the values of $\alpha$ that maximize the cosine with the shapes of different physical models. In figure \ref{app:fig:template} we show the form of the template that approximates the shape of the non-Gaussianity generated by the operator proportional to $M_3$.
\begin{table}[t]
 \begin{center}
 \begin{tabular}{|c|c|c|}
        \hline 
        Model &  $\alpha$ & $|\cos|$\\
        \hline         \hline 
	$M_3$ & 0.71 & 0.95 \\
	\hline
	\cite{Senatore:2009gt}, orth & 0.55 & 0.98 \\
        	\hline
        	\cite{Senatore:2009gt}, flat & 0.60 & 0.98 \\
	\hline
  \end{tabular}
  \end{center}
  \caption{\label{tab:alpha} \small Values of $\alpha$ maximizing the cosine of the template \eqref{app:eq:template} with different physical shapes, and corresponding value of the cosine. In the first line we give the values for $F_{M_3}$, in the second and third line we compare with the two shapes obtained in Ref. [17] as difference of equilateral shapes. Namely, the two shapes are obtained setting $\tilde{c}_3 = -5.4$ (orthogonal) and $\tilde{c}_3 = -6$ (flat) in Eq.~(16) of that Reference.}
\end{table}

\end{appendix}


\end{document}